\documentclass{optica-article}

\journal{opticajournal} 

\articletype{Research Article}

\usepackage{lineno}
\usepackage{subfigure}
\usepackage{placeins}

\begin{document}

\title{Wide-range resistivity characterization of semiconductors with terahertz time-domain spectroscopy}

\author{Joshua Hennig\authormark{1,2,*}, Jens Klier\authormark{1}, Stefan Duran\authormark{1}, Kuei-Shen Hsu\authormark{3}, Jan Beyer\authormark{3}, Christian Röder\authormark{4}, Franziska C. Beyer\authormark{4}, Nadine Schüler \authormark{5}, Nico Vieweg\authormark{6}, Katja Dutzi\authormark{6}, Georg von Freymann\authormark{1,2}, and Daniel Molter\authormark{1}}

\address{\authormark{1}Fraunhofer Institute for Industrial Mathematics ITWM, Department Materials Characterization and Testing, 67663\,Kaiserslautern, Germany\\
\authormark{2}Department of Physics and Research Center OPTIMAS, RPTU Kaiserslautern-Landau, 67663\,Kaiserslautern, Germany\\
\authormark{3}Institute of Applied Physics, Technische Universität Bergakademie Freiberg, Leipziger Str. 23, 09599\,Freiberg, Germany\\
\authormark{4}Fraunhofer Institute for Integrated Systems and Device Technology IISB, Department Energy Materials and Test Devices, Schottkystraße 10, 91058\,Erlangen, Germany\\
\authormark{5}Freiberg Instruments GmbH, Delfter Str. 6, 09599\,Freiberg, Germany\\
\authormark{6}TOPTICA Photonics AG, Lochhamer Schlag 19, 82166\,Gräfelfing, Germany\\}

\email{\authormark{*}joshua.hennig@itwm.fraunhofer.de} 


\begin{abstract*} 
Resistivity is one of the most important characteristics in the semiconductor industry. The most common way to measure resistivity is the four-point probe method, which requires physical contact with the material under test. Terahertz time domain spectroscopy, a fast and non-destructive measurement method, is already well established in the characterization of dielectrics. In this work, we demonstrate the potential of two Drude model-based approaches to extract resistivity values from terahertz time-domain spectroscopy measurements of silicon in a wide range from about 10$^{-3}$\,$\Omega$cm to 10$^{2}$\,$\Omega$cm. One method is an analytical approach and the other is an optimization approach. Four-point probe measurements are used as a reference. In addition, the spatial resistivity distribution is imaged by X-Y scanning of the samples to detect inhomogeneities in the doping distribution.

\end{abstract*}

\section{Introduction} 
In the modern semiconductor industry, the fast, accurate and non-destructive determination of resistivity and doping level, as well as their homogeneity, is very important for the quality control of materials. An established method for these measurements is the four-point probe (4PP) measurement. However, this is a contact-based method, which means that the material may be affected by the 4PP measurement. In addition, its measurement speed is limited by the movement of the probe tip. Furthermore, the 4PP measurement only allows to extract information from the surface of the sample \cite{Smits1958, Miccoli2015, Naftaly2021}.

Terahertz time domain spectroscopy (TDS) promises to overcome these drawbacks and therefore has great potential in the characterization of semiconductors because the plasma frequency of semiconductors, which depends on the doping, lies in the terahertz range. Since its introduction as a non-destructive measurement technique \cite{Cheung1986, VanExter1989}, TDS has become an established method for material characterization purposes \cite{Naftaly2007, Hangyo2005, Theuer2011, Jepsen2011}. Especially the extraction of material parameters such as thickness, refractive index and absorption is used in industry and science \cite{Ellrich2016, Cunningham2011}. 

For characterizing semiconductors with TDS \cite{Jeon1997}, mostly the transmission geometry has been used \cite{Grischkowsky1990, Katzenellenbogen1992, Zhang2003, Naftaly2007}. Due to the increasing terahertz absorption with increasing free carrier density, this approach is limited to either high resistivity wafers \cite{Dai2004} or very thin films \cite{Herrmann2002, Yasuda2008}. Measuring in reflection geometry has the advantage of allowing the measurement of low resistivity samples that would not be accessible in transmission. In addition, modern applications use two-layer systems consisting of an epitaxially grown layer on a substrate of either high or low resistivity, again making reflection the more universal approach. However, many of the previous studies using reflection measurements are limited to a small number of samples in a small resistivity range \cite{Naftaly2021, Jeon1998, Guo2009, Alberding2017}. The same can be said about the work on mapping whole wafers \cite{Hamano2014}. Together with recent advances in measurement rate \cite{Dietz2014} and a rotating measurement setup \cite{Molter2022}, the speed of this technique has been accelerated up to the kilohertz range. This shows that TDS can be a useful tool as a non-destructive and even non-contact method, e.g. for quality control and characterization in the production or further manufacturing process of semiconductor wafers. With the wide range of resistivities that can be distinguished by TDS, it may even become competitive with established techniques for resistivity characterization of silicon or other semiconductors. 

For silicon wafers, especially in the case of highly doped wafers with a therefore low resistivity, the Drude model can be applied to describe the optical and electrical properties \cite{Nashima2001}. By using this simple model as well as the Fresnel coefficients, the relevant parameters of resistivity and carrier concentration are extracted from the TDS measurements. In order to show this, a variety of p- and n-doped silicon wafers within a wide range of resistivities is investigated. Mapping the wafers in an X-Y-pattern with a resolution of 1\,mm allows to detect doping inhomogenities. Combining the Drude model evaluation and the mapping therefore allows to characterize a wide range of silicon wafers precisely. 

\section{Methods}

\subsection{Terahertz TDS measurement}

\begin{figure}[htb!]
\centering
    \centering\includegraphics[width=0.9\textwidth]
    {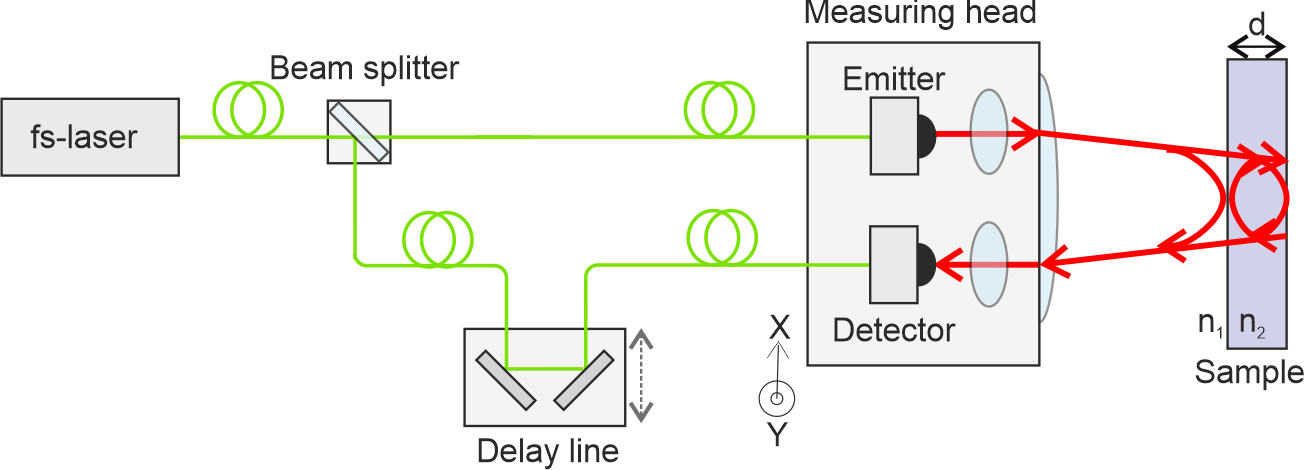}
    \caption{TDS schematic setup in reflection geometry. A fs-laser pulse is split by a beam splitter. One half of the pulse goes to the emitter emitting a terahertz beam. After the reflection measurement this terahertz beam is detected by a detector. By adjusting the optical path length difference with the delay line the terahertz pulse is sampled and finally detected completely.}
    \label{fig:TDS}
\end{figure}

The TDS system used in this work is a fiber-coupled setup in reflection geometry that is schematically shown in Fig. \ref{fig:TDS}. The setup is very similar to TDS systems used in many different applications, consisting of a fs-laser, a delay line and photoconductive antennas (PCA) serving as emitter and detector \cite{Weber2020, Castro-Camus2016}. The terahertz pulse coming from the emitting PCA is focused onto the sample of thickness $d$ and reflected into the detector. The angle between the incoming and outgoing beam is small (approx. 8 degrees). Hence, for the following considerations we assume normal incidence on the sample. The emitted terahertz pulses have a duration of a few picoseconds in the time domain and a frequency bandwidth of about 4\,THz. The focused spot size is roughly 1\,mm. 

\begin{figure}[htb!]
\centering
    \centering\includegraphics[width=0.9\textwidth]
    {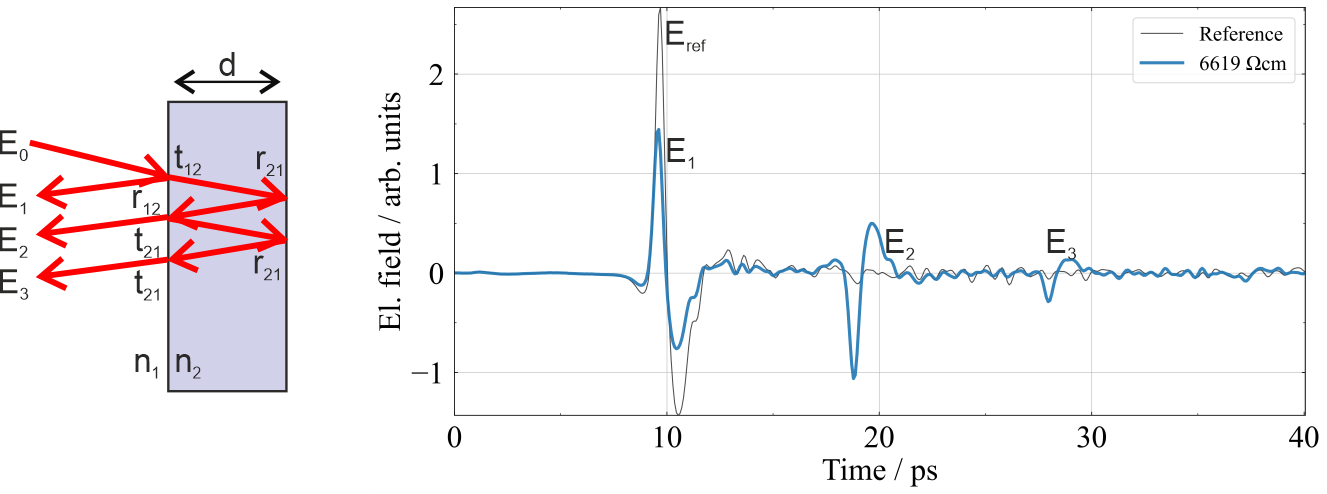}
    \put(-350,120){(a)}
    \put(-250,120){(b)}
    \caption{(a) The electric field of an incident terahertz beam E$_{0}$ is partially reflected from the front surface of a sample, leading to a first peak that can be detected E$_{1}$, and partially transmitted. The transmitted part undergoes multireflections inside the sample leading to electric fields E$_{2}$, E$_{3}$, respectively. At each surface between the surrounding air and the sample, corresponding Fresnel coefficients need to be considered. (b) A reference measurement as well as an example measurement in the time domain of a high-resistivity wafer showing the corresponding peaks.} 
   \label{fig:THz-sample_interaction+time_domain_2_measurements}   
\end{figure}

For a reflection measurement, the interaction of the terahertz pulse with the sample is sketched in Fig. \ref{fig:THz-sample_interaction+time_domain_2_measurements} (a). At each interface, the terahertz radiation is partially reflected and partially transmitted. For every measurement a reference measurement with an aluminium mirror placed in the same position as the sample is recorded first. Because the aluminium is highly reflective in the terahertz frequency range, the recorded electric field amplitude $E_{\text{ref}}$ can be considered as almost equal to the incident electric field $E_{\text{0}}$. Fig. \ref{fig:THz-sample_interaction+time_domain_2_measurements} (b) shows a reference measurement and one measurement of an n-doped silicon wafer with a high resistivity. In the measurement with a silicon wafer of thickness $d$, the first detected peak originates from the reflection at the front surface of the wafer. Therefore, the relative amplitude in the measurement equals the reflection coefficient $r_{12}$ between the ambient air and the corresponding silicon sample
\begin{equation}
    \frac{E_{1}}{E_{\text{ref}}} = r_{12}.
    \label{eq:peak_1_rel_amplitude}
\end{equation}
In an etalon-like beam path, the second peak and its electric field amplitude $E_{\text{2}}$ originate from terahertz radiation entering the sample, being reflected at the back surface between silicon and air and leaving the sample again. The shape of $E_{\text{2}}$ is inverted compared to $E_{\text{1}}$ and $E_{\text{ref}}$ as their reflection at the front surface causes a phase shift of $\pi$ because it is a reflection at an optically thicker medium. Additionally, for the distance $2d$ traversed inside the sample absorption according to the Beer-Lambert law leads to an exponential decrease in amplitude that is mathematically summarized by the extinction coefficient $\alpha$. This uses the corresponding transmission coefficients $t_{12}$ as well $t_{21}$ and is also under the approximation of normal incidence. Therefore, the relative field amplitude detected in the second peak can be described by
\begin{equation}
    \frac{E_{2}}{E_{\text{ref}}} = t_{12}r_{21}t_{21}e^{-\alpha2d}.
    \label{eq:peak_2_rel_amplitude}
\end{equation}
Analogously, another two reflections inside the sample add two additional reflection coefficients $r_{21}$ to the equation including another two passages through the thickness $d$ of the wafer leading to more absorption losses. So the total peak amplitude of the third detectable peak is given by
\begin{equation}
    \frac{E_{3}}{E_{\text{ref}}} = t_{12}r_{21}r_{21}r_{21}t_{21}e^{-\alpha4d}.
    \label{eq:peak_3_rel_amplitude}
\end{equation}
In this setup the linearly polarized terahertz radiation is oriented perpendicularly to the plane of incidence and has an almost normal incidence angle. So the Fresnel coefficients appearing in equations (\ref{eq:peak_1_rel_amplitude}), (\ref{eq:peak_2_rel_amplitude}), and (\ref{eq:peak_3_rel_amplitude}) can be approximately described by the Fresnel equations for light under normal incidence.
The reflection coefficient is given by

\begin{equation}
    r_{12} = \frac{n_{1}-n_{2}}{n_{1}+n_{2}},
    \label{eq:Fresnel_r}
\end{equation}
and the transmission coefficient
\begin{equation}
    t_{12} = \frac{2n_{1}}{n_{1}+n_{2}}.
    \label{eq:Fresnel_t}
\end{equation}

In the experimental realization, it is important to ensure the adjustment of the setup, specifically to keep the sample at the focal length of the focused terahertz beam within the Rayleigh length.

To map the resistivity, the measurement head is moved axially with an X-Y-translation stage across the whole area of the wafer in a pixel by pixel fashion. The step size of the mapping is freely variable. At the focal plane, the lateral spot size of the terahertz beam is about 1\,mm in diameter. So, in order to fully map the sample, a step size of 1\,mm both in X- and Y-direction is used in the measurements.  The measurements presented in this work are averaged 10 times at each position. With the system working at 40\,Hz, this leads to a measurement time at each position of 250\,ms. So, for a 4" wafer, the pure data acquisition time is around 53 minutes. However, also measuring about 5\,mm additionally at each edge in order to guarantee to measure the whole wafer and also taking the movement time of the axis into account, the real measurement time in that case is about 140 minutes with the described setup. Improving the measurement procedure and using a faster system is feasible \cite{Molter2022}, but outside the scope of the proof-of-concept nature of this work.

\subsection{Drude Model}
The Drude model, a simple model to describe electrical conduction in metals, has been proven to be suitable to describe the electrical properties of silicon, especially if it is highly doped and therefore has a low resistivity \cite{Nashima2001}. Although there are approaches to a more detailed description of the properties of silicon \cite{VanExter1990, Willis2013}, a distinction between different doping levels is already possible with the simple Drude model. The complex relative permittivity $\tilde{\epsilon}$ in this model is expressed by
\begin{equation}
    \tilde{\epsilon} = \epsilon_{\text{Si}}-\frac{\omega_{\text{p}}^{2}}{\omega(\omega+i\Gamma)}.
    \label{eq:dielectric_function_Drude}
\end{equation}
Here, $\epsilon_{\text{Si}} = (3.418)^{2} \approx 11.68$ is the dielectric constant of undoped silicon \cite{Dai2004}, $\omega_{\text{p}}^{2} = Nq^{2}/(\epsilon_{0}m_{\text{e}}m^{*})$ the plasma frequency squared, $\omega$ the angular frequency, $i$ the imaginary unit and $\Gamma = 1/\tau$ the damping frequency which is the inverse of the mean free time $\tau = \frac{m^{*}m_{\text{e}}\mu}{q}$. $N$ is the carrier density, $q$ the elementary charge, $\epsilon_{0}$ the vacuum permittivity, $m_{\text{e}}$ the electron mass and $m^{*}$ the effective mass for conductivity, either the effective mass of electrons $m_{\text{n}}^{*}=0.26$ in case of n-doping or the effective mass of holes $m_{\text{p}}^{*}=0.37$ in case of p-doping \cite{Spitzer1957, Jeon1997}. 

For silicon the charge carrier mobility $\mu$ can be well described as a function of $N$  by the empirical equation 
\begin{equation}
    \mu(N) = \frac{\mu_{\text{max}} - \mu_{\text{min}}}{ 1 + \big (\frac{N}{N_{\text{ref}}} \big )^{\beta}} + \mu_{\text{min}}
    \label{eq:mobility_empirical_formula}
\end{equation}
with the material specific constant properties $\mu_{\text{max}}$, $\mu_{\text{min}}$, $N_{\text{ref}}$ and $\beta$ taken from \cite{Caughey1967} in the following, distinguishing the two cases of n-doping and p-doping, respectively. The exponent $\beta$, which is named $\alpha$ in \cite{Caughey1967}, is renamed in order to avoid confusion of the exponent with the extinction coefficient $\alpha$. Beside that the naming is the same. Substituting the plasma frequency and the damping frequency in equation (\ref{eq:dielectric_function_Drude}) and using the mentioned values for the other quantities leads to a formula for $\tilde{\epsilon}$ only depending on the charge carrier density and the angular frequency 
\begin{equation}
    \tilde{\epsilon}(N, \omega) = \epsilon_{\text{Si}}-\frac{Nq^{2}}{\epsilon_{0}m_{\text{e}}m^{*}\omega \big (\omega+i\frac{q}{m_{\text{e}}m^{*}\mu(N)} \big )}.
    \label{eq:dielectric_function_Drude_N_dependend}
\end{equation}
The resistivity $\rho$ can be calculated from the charge carrier density and the mobility via
\begin{equation}
    \rho  = \frac{1}{qN\mu(N)}.
    \label{eq:rho_formula}
\end{equation}
This is the desired quantity in this evaluation which will be compared to the resistivity values measured by the well established four-point probe (4PP) method serving as a reference method.

\subsection{Analytical evaluation method}
\begin{figure}[htb!]
\centering
    \centering\includegraphics[width=0.9\textwidth]{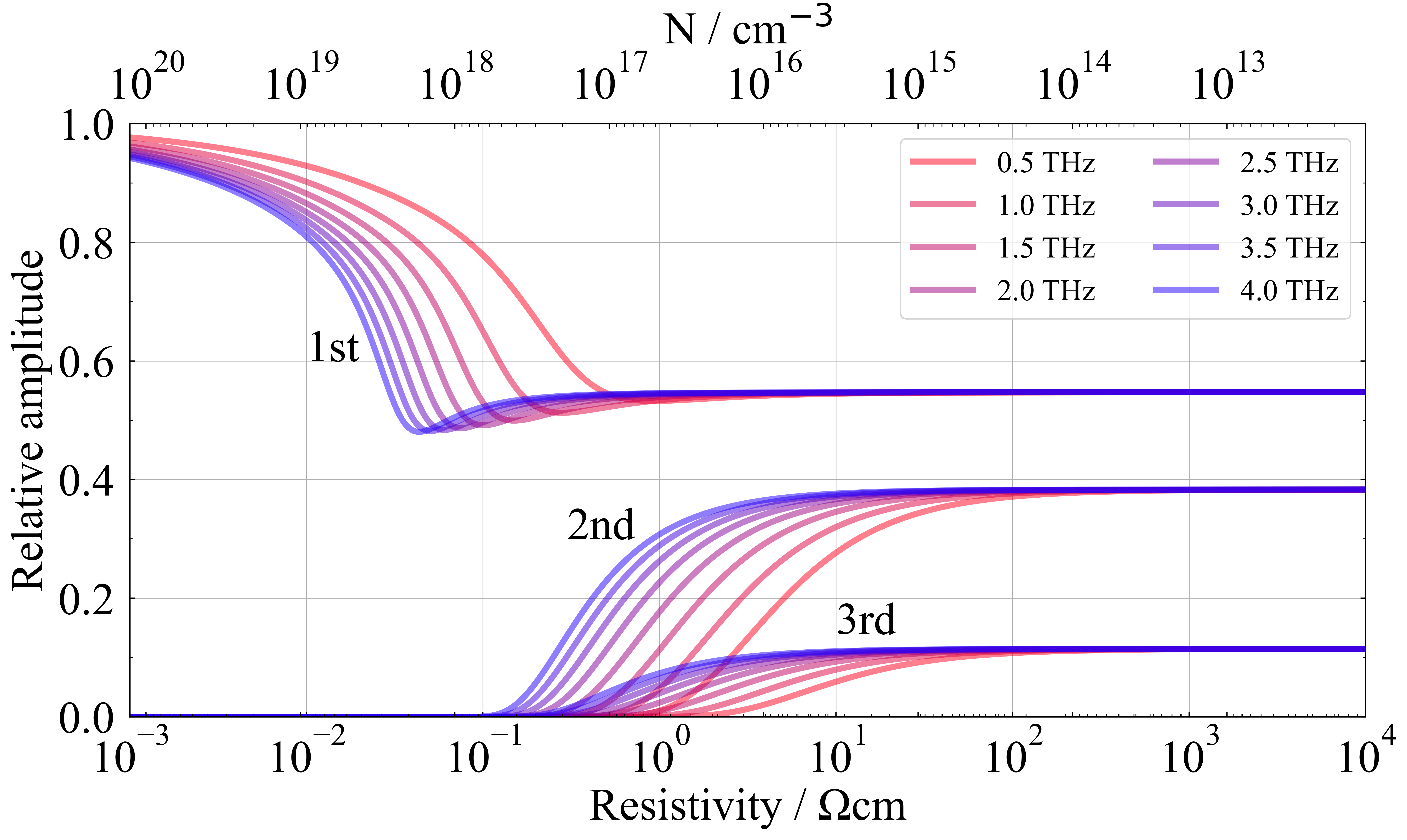}
    \caption{Curves of the theoretically expected relative peak amplitude of the 1st, 2nd and 3rd peak depending on the resistivity and the corresponding charge carrier density for different frequencies for the first three observable peaks by TDS in reflection geometry of p-type silicon according to the Drude model. A thickness of 400\,$\mu$m is assumed for these calculations.}
    \label{fig:Drude_model_rel_Amplitude}
\end{figure}

In the first part of the following evaluation, equation (\ref{eq:dielectric_function_Drude_N_dependend}) is used to calculate a look-up table for a doping-dependent relative permittivity. This is computationally easier to achieve than to rearrange the formula due to it including complex values as well as multiple N-dependencies. For this purpose, the frequency is assumed to be 0.5\,THz, which is roughly the median frequency of the used terahertz pulses in the setup. The complex refractive index $\tilde{n}$ can be calculated from this value using the equations for its real part 
\begin{equation}
    n_{r} = \frac{1}{\sqrt{2}}\sqrt{\epsilon_{i}+(\epsilon_{i}^{2}+\epsilon_{r}^{2})^\frac{1}{2}}
    \label{eq:eps_to_n}
\end{equation}
and its imaginary part
\begin{equation}
    n_{i} = \frac{1}{\sqrt{2}}\sqrt{-\epsilon_{i}+(\epsilon_{i}^{2}+\epsilon_{r}^{2})^\frac{1}{2}}.
    \label{eq:eps_to_k}
\end{equation}
In equations (\ref{eq:eps_to_n}) and (\ref{eq:eps_to_k}) $\epsilon_{r}$ and $\epsilon_{i}$ are the real and imaginary part of the relative permittivity in equation (\ref{eq:dielectric_function_Drude_N_dependend}), respectively. The obtained refractive index can then be used to calculate the corresponding Fresnel coefficients. For this purpose, the equations (\ref{eq:Fresnel_r}) and (\ref{eq:Fresnel_t}) and the complex refractive index of air in the terahertz range, which is assumed to be $n_{\text{air}} = 1.00027 + 0i$ \cite{Dai2004}, are used.
With the Fresnel coefficients and the thickness value, that is either given by the manufacturer information of each wafer or alternatively measured with a commercial thickness gauge, equations (\ref{eq:peak_1_rel_amplitude}), (\ref{eq:peak_2_rel_amplitude}) and (\ref{eq:peak_3_rel_amplitude}) are used to calculate the relative amplitude of the corresponding peaks. This allows for a direct correlation of the measured peak amplitude to the doping level of the silicon sample as well as the other quantities mentioned above.

By using this correspondence, the series of curves shown in Fig. \ref{fig:Drude_model_rel_Amplitude} can be obtained as the expected theoretical curves of the relative amplitude depending on the resistivity or the charge carrier density. This is exemplarily shown for p-doped silicon, and shows a similar behavior for n-doped silicon. The three groups of curves correspond to the three observed reflection peaks. The different colors of the curves indicate the frequency dependence. So, each frequency component in a measurement has a slightly different behavior, which can be taken advantage of in the optimization evaluation method. 

For resistivities below 0.5\,$\Omega$cm only one peak with a high amplitude is observable. For the middle range of resistivities around 1 $\Omega$cm a second and even a third peak appear. Due to the Fresnel losses, they have a much smaller amplitude than the first peak, of which the amplitude becomes constant here. The amplitudes of the second and third peak rise with increasing resistivity up to around 10$^{2}$\,$\Omega$cm where they saturate. This leads to a predicted sensitivity range of the TDS resistivity determination spanning across five orders of magnitude from around 10$^{-3}$\,$\Omega$cm to 10$^{2}$\,$\Omega$cm.

\subsection{Optimization evaluation method}\label{sect:method:optim}
The previously described analytical method has the drawback that it only allows one quantity to be variable and therefore has to rely on the thickness information available and only takes the median frequency into account whereas the pulse used in the experiments in fact has a bandwidth of around 4\,THz. In order to account for the dispersion of the material in the investigated frequency range up to about 4\,THz, an approach of simulating the pulse interacting with the material and optimizing it consecutively is taken. For this evaluation, the numeric Rouard's method is used to model the interaction with the silicon under investigation \cite{Krimi2016Diss, Vasicek1950}. The transfer function in this case of a single layer of silicon with air both in front and behind is given by
\begin{equation}
    H(\omega) = r_{\text{12}} + \frac{t_{\text{12}} r_{\text{21}} t_{\text{21}}\text{exp} \big (\frac{2i\tilde{n}\omega d}{c} \big )}{1-r_{\text{21}}r_{\text{21}}\text{exp} \big (\frac{2i\tilde{n}\omega d}{c} \big )}.
    \label{eq:Rouard_transfer_function}
\end{equation}
In order to calculate the Fresnel coefficients, the Drude model is utilized in the way mentioned before. Based on the Fourier transform of a reference measurement, that is done on an aluminium mirror, the pulse interacting with the material is simulated in the frequency domain. By the inverse Fast Fourier transform it is transferred back into the time domain and its waveform is then compared to the one measured \cite{Krimi2016}. In addition to making use of the whole frequency spectrum, this approach can take thickness variations into account by not only optimizing for the resistivity, but also optimizing the thickness value $d$ in certain boundaries.

In the evaluation, an optimization of the simulated waveforms is executed. Therefore, boundaries of $N$ from $10^{12}$\,$\Omega$cm to $10^{20}$\,$\Omega$cm and for the transparent samples boundaries for the thickness values in the range of the manufacturer given thickness $\pm 25$\,$\mu$m are used. These boundaries span the whole range of possible values of $N$ as well as a sufficient range to allow for thickness variations of the wafers. The function $f(R^{2}) = 1-R^{2}$ with the correlation coefficient $R$ between each of the simulations and the measurement serves as a function to be minimized. From the resulting value of N,  the resistivity $\rho$ can be calculated using equation (\ref{eq:rho_formula}).

This method has the potential to integrate even more parameters into the optimization. It can therefore potentially account for even more variations between materials and samples not considered yet, such as the effective mass or even multilayer samples.

\subsection{4PP method}
The four-point probe measurement method has been a well-established industrial standard method for determining sheet resistance since the American Society for Testing and Materials has recognized it as a reference procedure in 1975. The benefit of using the 4PP method is that the contribution of contact resistance between the probe and the surface is minimized by an in-line four-point setup. The probe spacing between each probe is 500\,µm. A current $I$ is injected by the outer two probes while the inner two probes measure the voltage drop $V$ caused by the current flow. The impedance of the voltage probe is considered as infinite, therefore the sample sheet resistance $R_{\square}$ can be obtained by the ratio of $V/I$. Additionally, a thickness-dependent correction factor is applied. The measurement range for $R_{\square}$ of the 4PP system used in this work (Model 280SI, Four Dimensions Inc.) ranges from $10^{-3}$\,$\frac{\Omega}{\square}$ to $2\times10^{5}$\,$\frac{\Omega}{\square}$. By convention, the unit of the sheet resistance is $\frac{\Omega}{\square}$ with the square ($\square$) being dimensionless, only representing the fact that it is a sheet resistance and not a bulk resistance. From the measured sheet resistance, the resistivity of each sample is calculated by multiplying $R_{\square}$ with the thickness $d$ of the sample.

\section{Measurement and results}
\subsection{Samples}
\begin{figure}[htb!]
\centering
    \centering\includegraphics[width=0.9\textwidth]{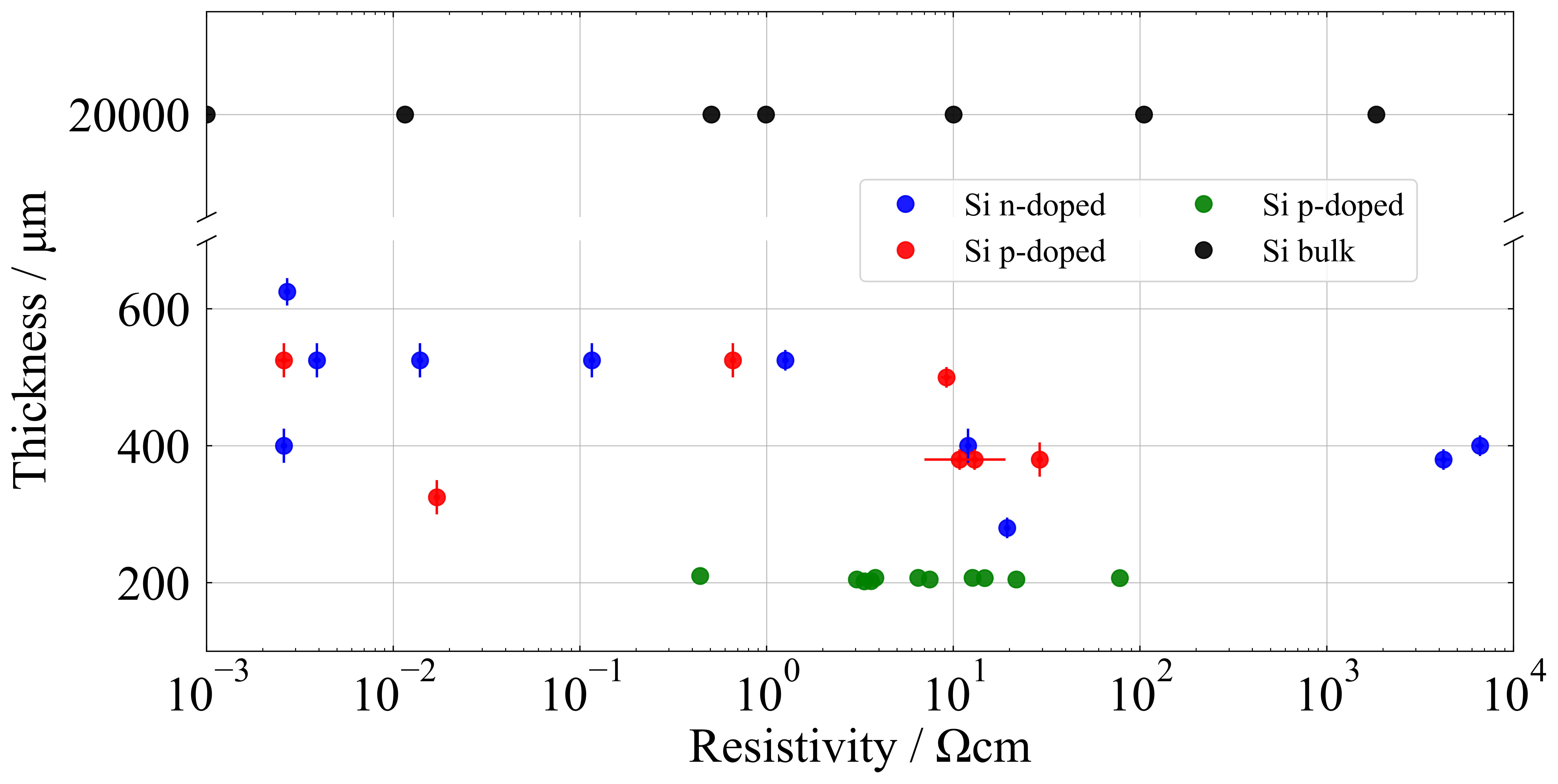}
    \caption{Overview of the samples investigated and their thicknesses and resistivities. The resistivity values are the mean of either a full mapping with a resolution of 8\,mm (blue and red markers) or a line scan (green and black markers) of 
    4PP-reference measurements across the whole wafer. The thickness values are either from the manufacturer's information (n-doped (blue markers), p-doped (red markers) and bulk (black markers), or measured with a commercial thickness gauge with a specified accuracy of 1\,$\mu$m (p-doped samples (green markers)).}
    \label{fig:rho_d_plot}
\end{figure}

With our setup and this theoretical approach, four sets of silicon samples were investigated. One of them consists of 10 n-doped wafers with thicknesses around 400\,$\mu$m, one consists of 9 p-doped wafers with thicknesses around 400\,$\mu$m, one consists of 11 p-doped wafers with thicknesses around 200\,$\mu$m, and one is a set of 7 bulk silicon samples with thicknesses around 20\,mm. Fig.~\ref{fig:rho_d_plot} shows the thickness and the resistivities that are characterized as the mean of a wafer-size dependent number of 4PP-measurements conducted across the whole wafer as a reference of the wafers investigated. For the n-doped and the red-marked p-doped sample sets, this reference is based on full maps of 4PP-measurements with spatial resolutions of about 8\,mm independent of the size of the wafers leading to 25 to 289 data points. For the green-marked p-doped set of wafers as well as the samples of bulk silicon only the mean value of an initial line-scan characterization with 5 data points is available making these values less comparable, especially concerning the mapping. The wafers used are of sizes in between 2" and 6" with a few of them being fragmented into multiple parts. 

Terahertz TDS imaging is performed by scanning the focused terahertz beam across the horizontal wafer surface in an X-Y raster scan fashion described above. Directly before the measurements on each wafer, a reference measurement on an aluminium mirror is recorded with the same system setup to make sure that the corresponding reference is measured under the most similar ambient temperature and humidity circumstances possible. The focal distance is adjusted with an adjustment screw at the measuring head. In a first step to do so, the maximum amplitude on the reference is adjusted and the live signal captured in the time domain. Then the wafer is placed  in the setup and adjusted to the same position in time leading to both the reference and the sample surface being placed at the same position in the focal point. To make sure the measurement is not affected by possible reflections from the metal of the optical table, that the axes are mounted on, a wafer holder fabricated for this purpose is used. It consists of a ground plate, that can be mounted on the table, and three thin adjustable pillar-like holders positioned in a triangular arrangement, each with a plastic tip. This way, the height of each of the three pillars can be adjusted separately allowing to place the samples as horizontally as possible, which is checked by moving the measurement head across the whole area of the sample, keeping the terahertz pulse reflection at the same time position. This way, the assumption that the terahertz radiation is focused on the sample under normal incidence at every position of the wafer is reasonably justified. Due to the height of the pillars, any reflections from the table or ground plate of the holder do not affect the measurement because they are out of the time window of the measurement.

\subsection{Analytical evaluation}
\begin{figure}[htb!]
\centering
    \centering\includegraphics[width=0.9\textwidth]{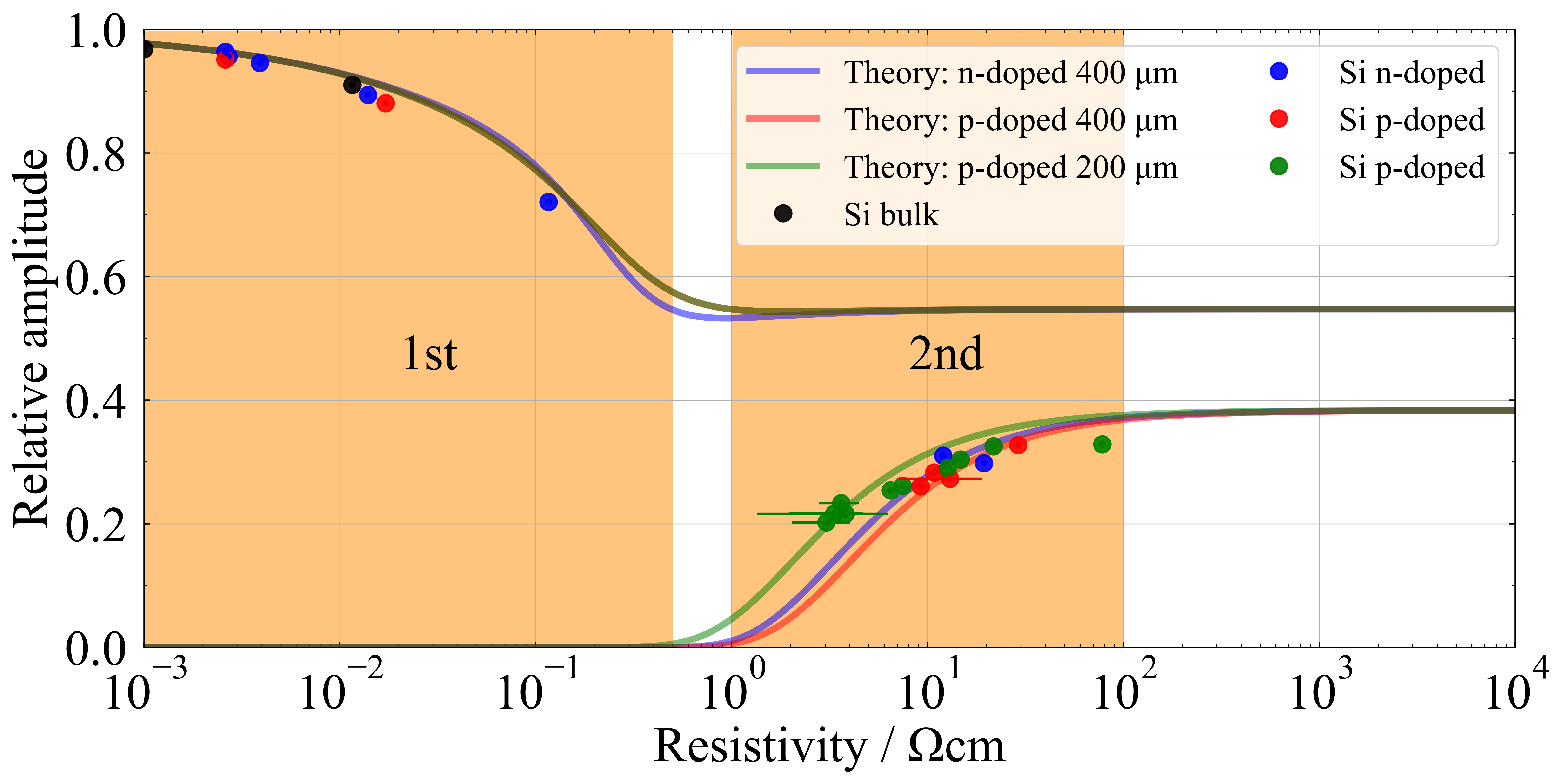}
    \caption{Resistivity-relative amplitude graph with theoretical curves for a median frequency of 0.5\,THz and measurement data of the relative amplitudes of the 1st and 2nd terahertz reflection peak in the time domain. The data points are at the position of the mean of approx. 7\% of the area of the wafers at their center. The standard deviation represented by the error bars is for most measurements smaller than the data points. The orange marked areas indicate in which resistivity range the 1st or 2nd terahertz reflection peak is used for the evaluation. Between about 0.5\,$\Omega$cm and 2\,$\Omega$cm there is a gap due to weak sensitivity in the peak amplitudes.}
    \label{fig:amplitude_rho_plot}
\end{figure}

The analytical evaluation method for the wafer resistivity is based on the relative amplitudes of the individual terahertz reflection peaks of the time domain measurement. Therefore, Fig. \ref{fig:amplitude_rho_plot} shows the three theoretical curves corresponding to each of the wafer sets with their corresponding doping and thickness. The plotted amplitude value of each wafer is the mean of a size dependent number of measurements in a centered area covering roughly 7\% of the area of each wafer (between 132 and 1404 measurements). The resistivity value plotted here is the reference resistivity value from the 4PP measurements. The error bars indicating the standard deviation are smaller than the data points for many of the samples. For each wafer under test, only the amplitude of the peak that is used in the next step of the evaluation is shown. For the low resistivity samples, only the first peak is used and for the higher resistivity wafers only the second peak is used. Because of the higher absolute value of the amplitude of the second peak compared to the third one, this one seems more promising to lead to more accurate results. Samples in the resistivity range of around 0.5\,$\Omega$cm to 2\,$\Omega$cm cannot be evaluated with this method because the dynamic in the amplitude behavior is too low. The same holds true for samples with resistivities above 100\,$\Omega$cm. Therefore, all samples with resistivities in these ranges are neglected here. Still, this method covers large parts of the resistivity range of high interest, given that e.g. typical doping levels of majority carriers in semiconductors in the field of electronics are often in the range of $10^{15}$~-~$10^{18}~$atoms/cm$^3$ \cite{Razavi} - see Fig. \ref{fig:Drude_model_rel_Amplitude} for the correspondence of the resistivity and the doping level ranges.

\begin{figure}[htb!]
\centering
    \centering\includegraphics[width=0.9\textwidth]
    {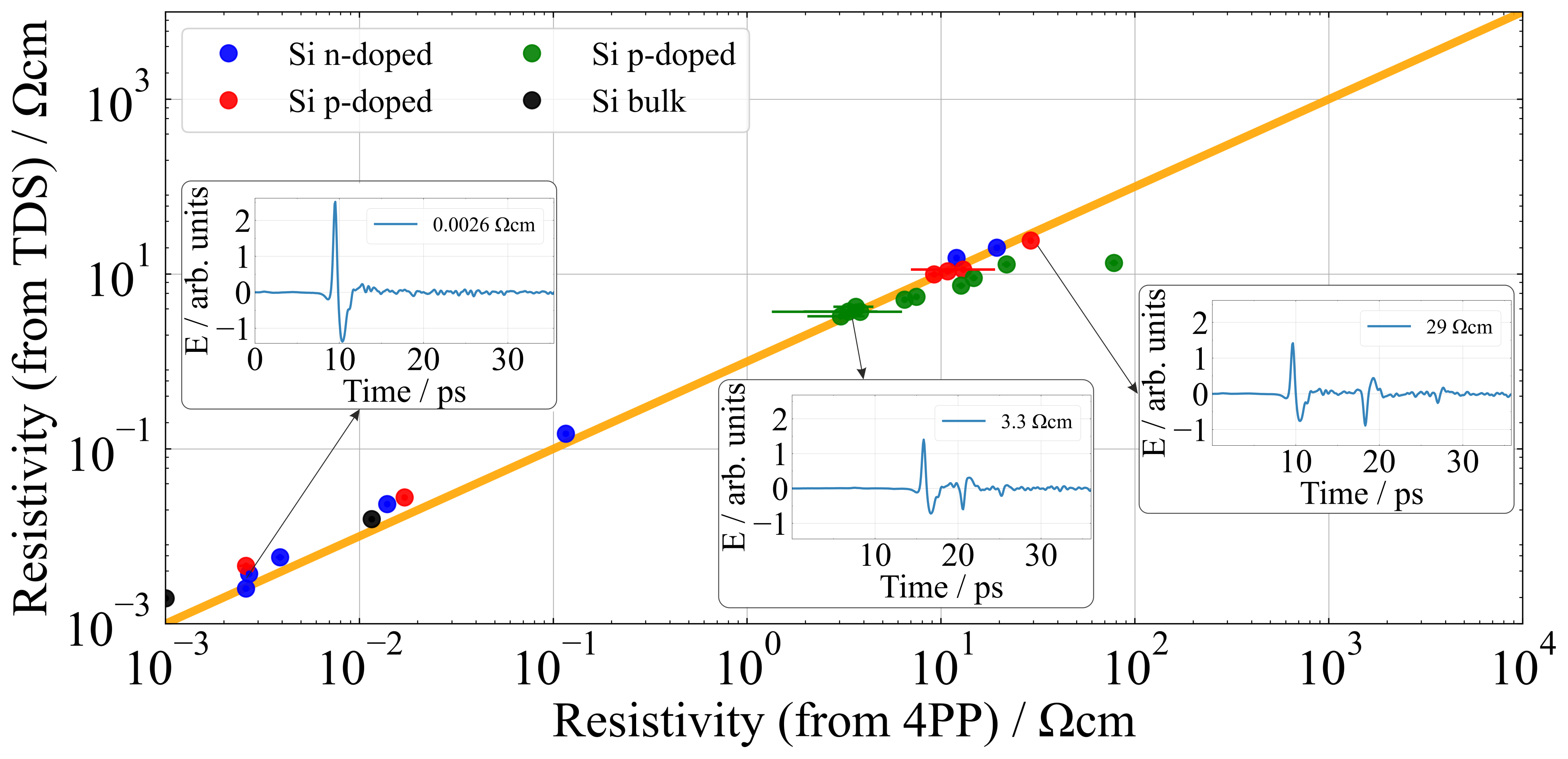}
    \caption{Resistivity obtained by the analytical evaluation method utilizing the Drude model depending on the 4PP-resistivity value. Shown are the mean and the standard deviation as the error bars of an area of approx. 7\% of each wafer. The orange curve is the angle bisector for eye guidance. In the insets three time domain measurements of samples with high, middle and low resistivity as well as different thicknesses are exemplarily shown.}
    \label{fig:rho_rho_analytical}
\end{figure}

The resistivity can be determined from the measured relative amplitudes as explained in the theory above. Fig. \ref{fig:rho_rho_analytical} shows the resistivity value from this evaluation compared to the one from the four-point probe measurement. The error bars are mostly smaller than the data points. It should be noted that both the terahertz TDS method and the 4PP method have certain errors or might not be calibrated or aligned perfectly, which could lead to small deviations. While for the 4PP measurement no metal deposition and thermal contact formation on the wafer is required, there are some potential errors such as dislocation in the sample by the probe, wear or oxidation of the probe tips or a nonlinearity in the contact resistance. For the TDS measurement, possible sources of error are slight defocusing or a small tilt of the samples. As mentioned before, both these errors are minimized by checking and aligning the setup before each measurement. Furthermore, variation in the ambient temperature as well as the air humidity could have an impact on the measurements, which is minimized by taking separate reference measurements for each sample.

However, for the resistivity range of around roughly 10$^{-3}$\,$\Omega$cm to 0.5\,$\Omega$cm the coefficient of determination of the obtained resistivity values R$^{2}$ equals 0.997. For the resistivity range from around roughly 2\,$\Omega$cm to  10$^{2}$\,$\Omega$cm R$^{2}$ equals 0.64 for the green-marked samples with only one mean reference four-point-probe value available. This comparably low coefficient of determination might be caused by the fact that it is not documented at which position on the wafer the line-scan reference measurement was done. R$^{2}$ for the other sets of samples in this resistivity range equals 0.89.

In conclusion, this method works well with good coefficients of determination in the two above-mentioned resistivity ranges. The drawback of a small gap in the resistivity axis in this evaluation method is obvious and it is addressed by the optimization method.

\FloatBarrier

\subsection{Optimization evaluation}
The method of simulating and optimizing the terahertz pulse and its interaction with the sample under test, which was explained in section \ref{sect:method:optim}, is on the one hand a more complicated and especially computationally more intense approach. But the benefit of using the whole spectral information of the measurement on the other hand promises a more realistic and complete result. 

\begin{figure}[htb!]
\centering
    \centering\includegraphics[width=0.9\textwidth]{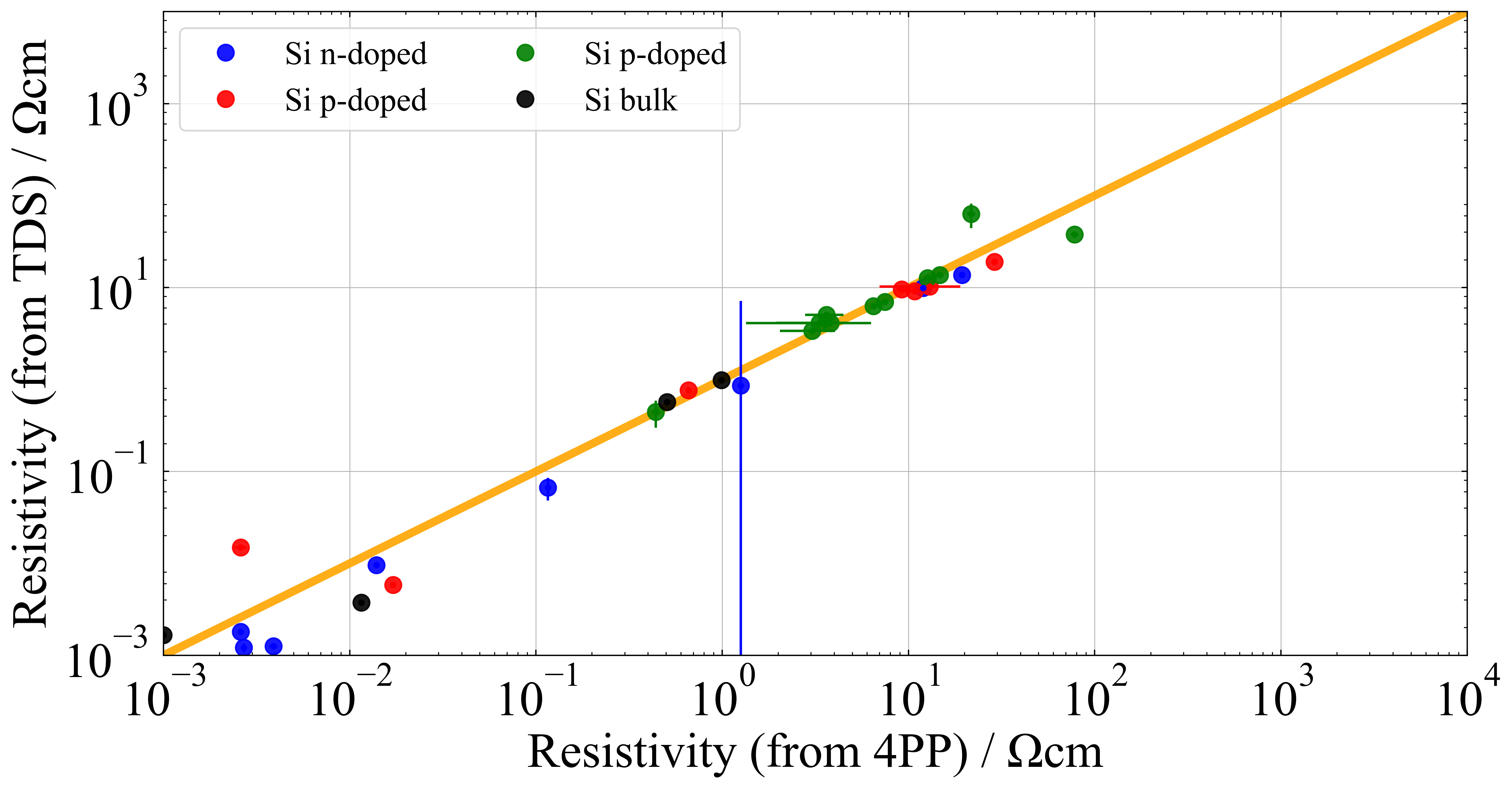}
    \caption{Resistivity obtained by the simulation evaluation method utilizing the Drude model depending on the 4PP-resistivity value. Shown are the mean and the standard deviation as the error bars of an area of approx. 7\% of each wafer. The orange curve is the angle bisector for eye guidance.}
    \label{fig:rho_rho_optimization}
\end{figure}

Fig. \ref{fig:rho_rho_optimization} shows the resulting resistivity values of this method compared to the 4PP reference values analogously to Fig. \ref{fig:rho_rho_analytical}. In the range of higher resistivities from 2\,$\Omega$cm to  10$^{2}$\,$\Omega$cm the coefficient of determination R$^{2}$ equals 0.36 for the green-marked p-doped sample set and R$^2$ equals 0.98 for the other sets. For resistivities between 0.5\,$\Omega$cm and 2\,$\Omega$cm, R$^{2}$ equals 0.70. So this method additionally covers this range of resistivities. The blue marked n-doped wafer with a 4PP-measured resistivity of around 1.26\,$\Omega$cm has a comparably high standard deviation. This is due to artefacts in the measurements. Therefore, this is the only sample, for which the boundaries of allowed values for N had to be reduced to the range from $10^{12}$\,cm$^{-3}$ to $10^{17}$\,cm$^{-3}$. This is reasonable because from the peak amplitude one can already estimate the range of N, see Fig. \ref{fig:Drude_model_rel_Amplitude}. However, for resistivities from 10$^{-3}$\,$\Omega$cm to  0.5\,$\Omega$cm R$^{2}$ equals 0.94.

A source of error could be the fact that the information from inside the sample is missing for low-resistivity samples and therefore only the surface is directly investigated by the reflection of the terahertz pulse. 
Also, the simulated pulse in the time domain, which is compared to the measured pulse over the whole time window, is only influenced in a small part of this time window, more precisely only at the position of the pulse. So, the rest of the measurement time that includes sources of error such as water oscillations or noise, can have a relatively higher impact on the overall result. This impact is much smaller when there is at least a second peak in the waveform extending the time of the measurement holding relevant material information of the sample. Therefore, the analytical result might benefit in the case of low resistivity from a lower statistical noise.

\subsection{Imaging and resistivity distribution}
So far, only averaged values for each sample have been considered. To make use of the full potential of the X-Y-scans of the wafers, mapping the results and thereby imaging the wafer and its resistivity distribution reveals as much information as possible from this analysis. In order to allow for a good comparison with the reference method, the four-point probe measurements of the samples are done in a scanning mode as well. However, the resolution of the 4PP-scans is limited to the maximum number of 625 measurements possible with the system, leading to a resolution of 3.6\,mm for a 4" wafer.
The more adaptable X-Y-scan of the TDS measurement has a higher resolution of 1\,mm. To make the two imaging techniques more comparable, the 4PP-measurements are interpolated using the nearest-neighbor method, which makes the flat vanish in the plot.

\begin{figure}[htb!]
\centering
    \begin{subfigure}{} 
        \includegraphics[width=0.4\textwidth]{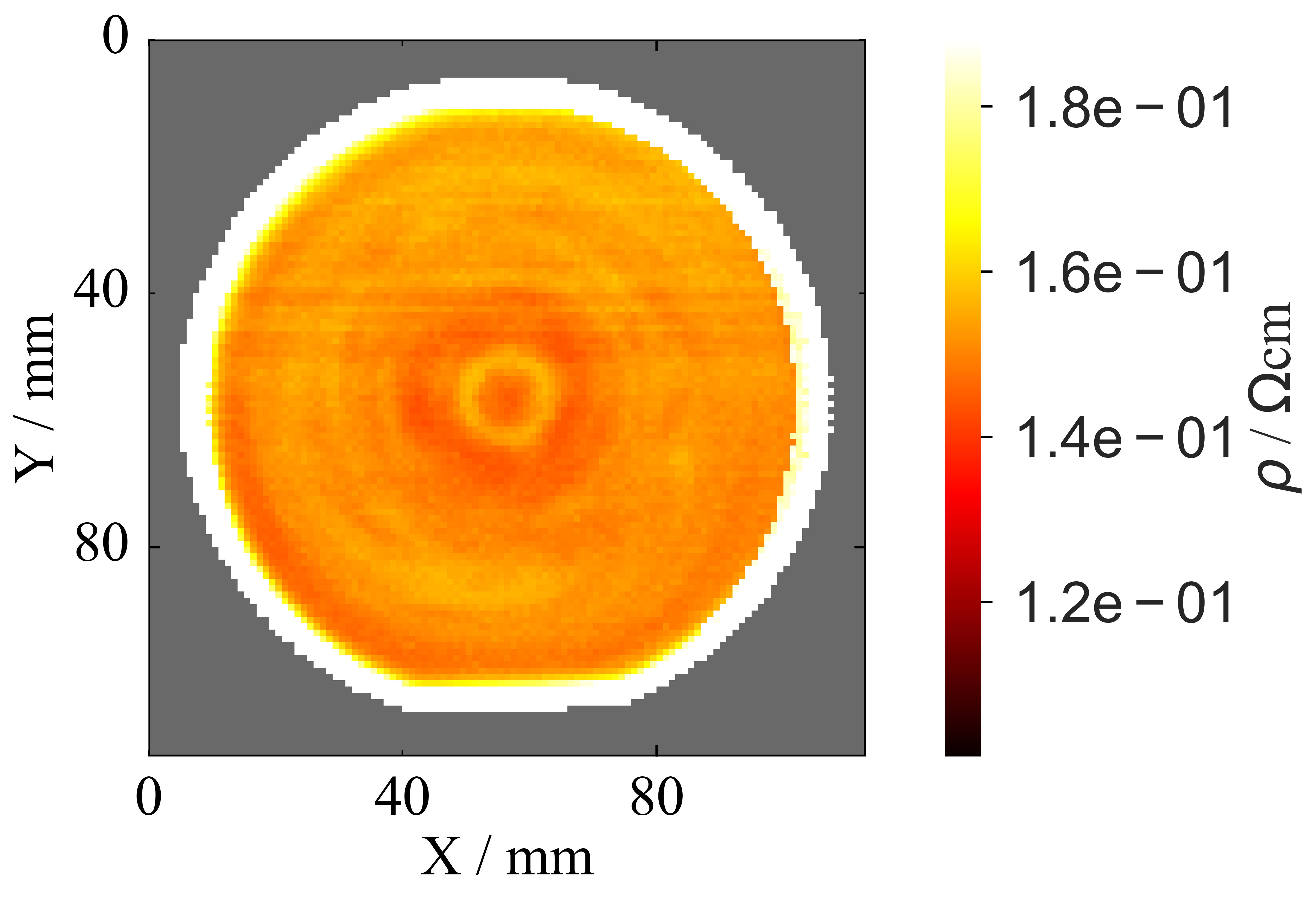}  
        \put(-155,95){(a)}
        \label{fig:Comparison_Imaging_TDS_4PP_structure_TDS}
    \end{subfigure}
    \begin{subfigure}{}
        \includegraphics[width=0.4\textwidth]{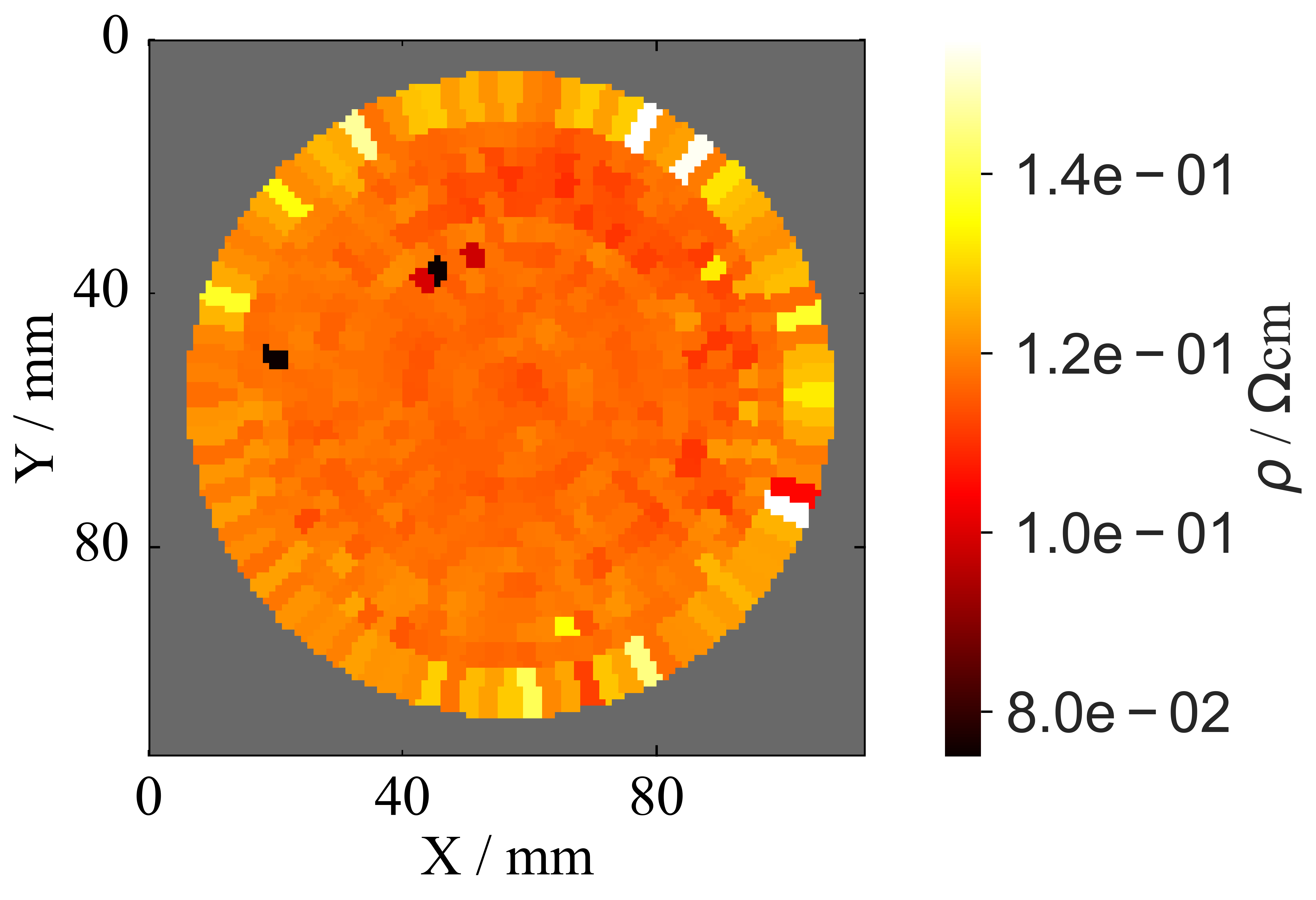}
        \put(-155,95){(b)}
        \label{fig:Comparison_Imaging_TDS_4PP_structure_4pp}
    \end{subfigure}
    \caption[]
    {\small Imaging results of the resistivity of the same n-doped wafer obtained by (a) the analytical evaluation method of the TDS scan measurement with a mean of the TDS-determined resistivity of 1.4$\times$10$^{-1}$\,$\Omega$cm and (b) the 4PP-mapping with a mean of the 4PP-measured resistivity of 1.1$\times$10$^{-1}$\,$\Omega$cm.}
    \label{fig:Comparison_Imaging_TDS_4PP_structure}    
\end{figure}

In Fig. \ref{fig:Comparison_Imaging_TDS_4PP_structure} (a) the calculated resistivity of the scan of the analytical evaluation method of the TDS measurement is compared to the resistivity obtained by (b) the scan of the 4PP measurement on the same medium-resistivity wafer. Ring-like resistivity striations, originating from variations in the dopant segregation coefficient, can be observed, which is a structure that is typical for silicon wafers produced by the Czochralski method \cite{Dornberger1996}.
Due to the better resolution of the scan, the structure is clearly observable in the TDS measurement. In the 4PP measurement the ring-like structure can be detected in faint outlines as well. The colorbar between (a) and (b) spans across the same resistivity range, but it is shifted by the offset of the mean of both methods for better comparability of the lateral resistivity distribution.

\begin{figure}[htb!]
\centering
    \begin{subfigure}{} 
        \includegraphics[width=0.4\textwidth]{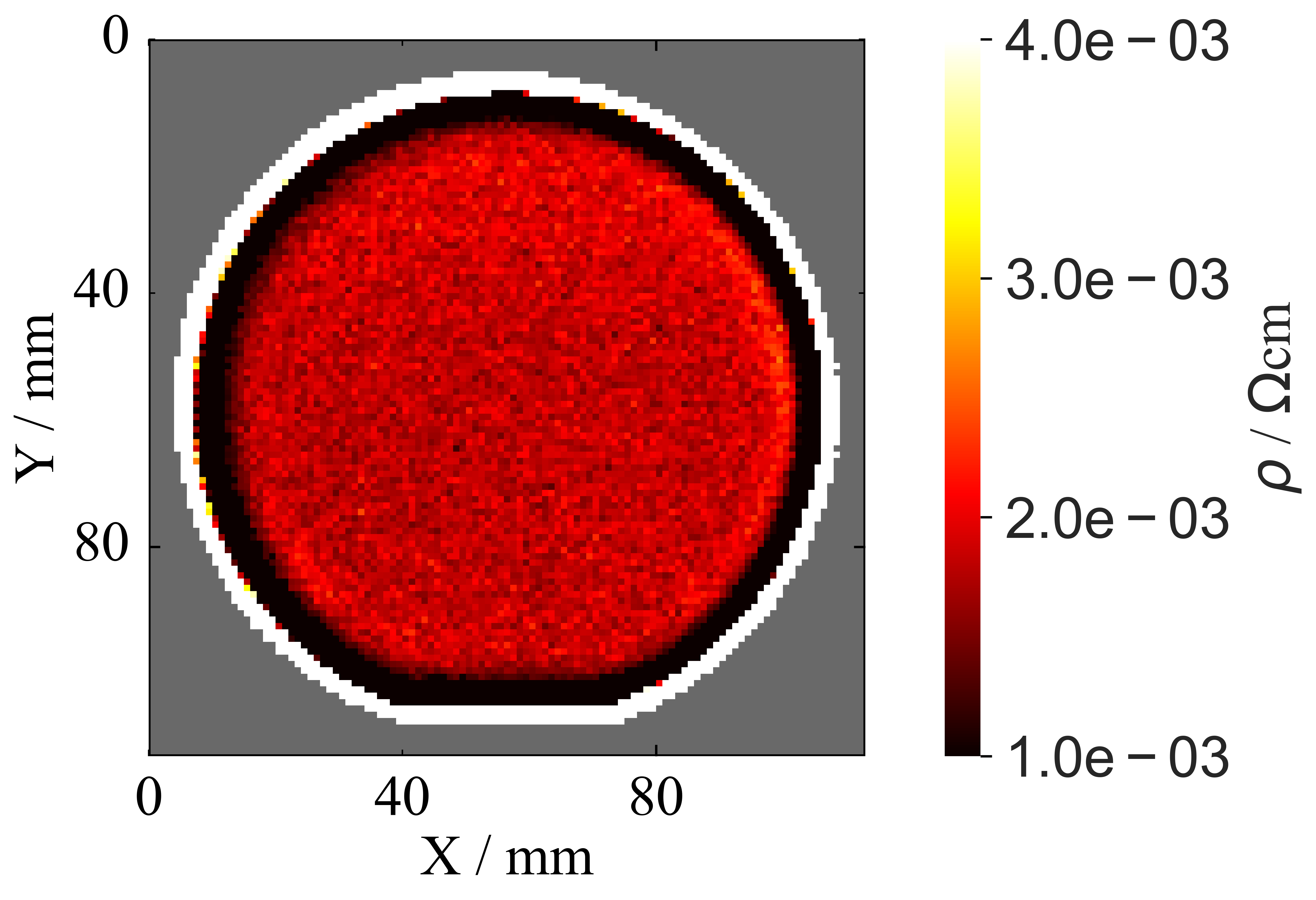}  
        \put(-155,95){(a)}
        \label{fig:Comparison_Imaging_TDS_4PP_TDS}
    \end{subfigure}
    \begin{subfigure}{}
        \includegraphics[width=0.4\textwidth]{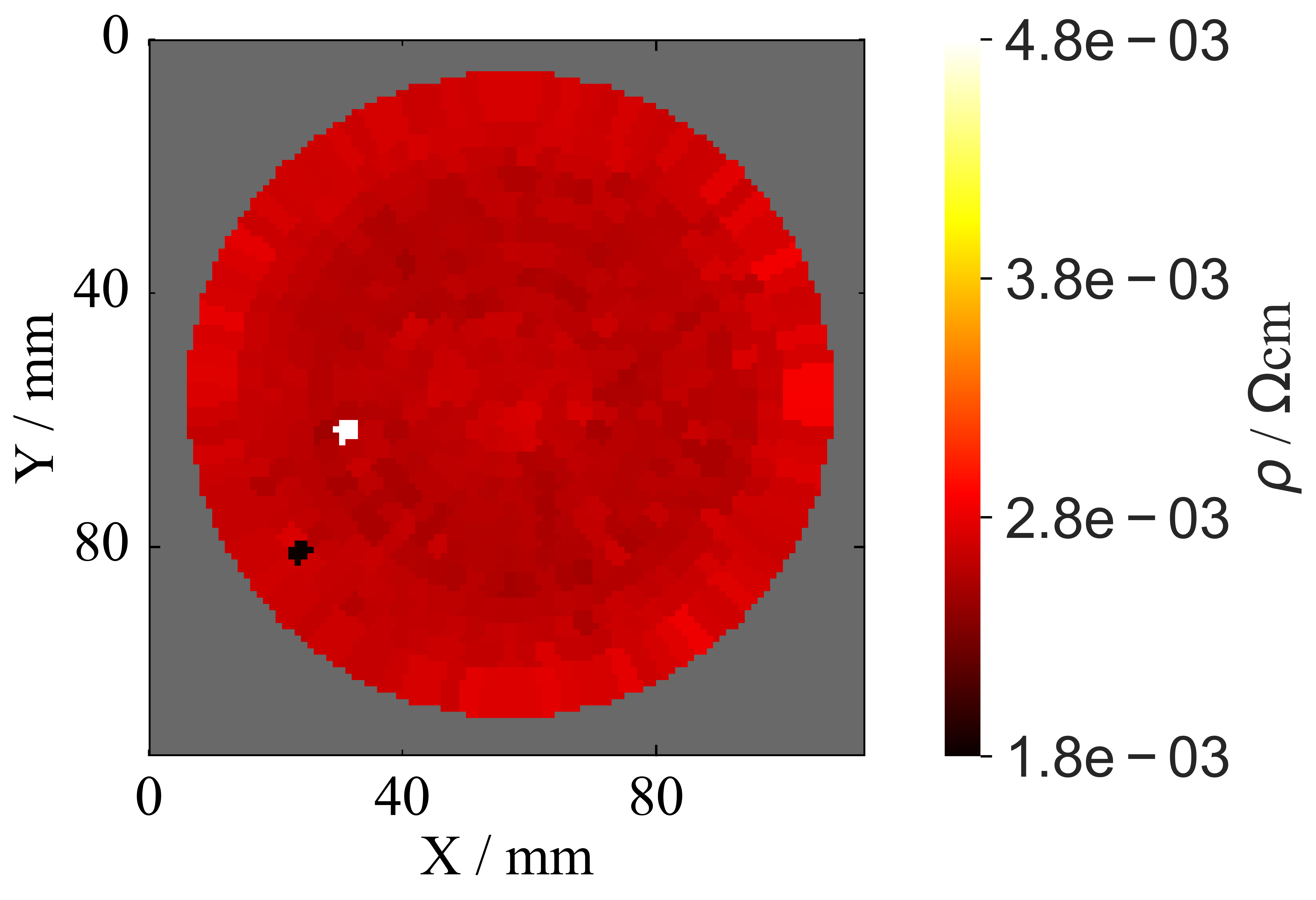}
        \put(-155,95){(b)}
        \label{fig:Comparison_Imaging_TDS_4PP_4pp}
    \end{subfigure}
    \caption[]
    {\small Imaging results of the resistivity of the same n-doped wafer obtained by (a) the optimization method evaluation of the TDS scan measurement with a mean of the TDS-determined resistivity of 1.8$\times$10$^{-3}$\,$\Omega$cm and (b) the 4PP-mapping with a mean of the 4PP-measured resistivity of 2.6$\times$10$^{-3}$\,$\Omega$cm.}
    \label{fig:Comparison_Imaging_TDS_4PP}    
\end{figure}

Fig. \ref{fig:Comparison_Imaging_TDS_4PP} shows a comparison of the resistivity maps of a low-resistivity wafer obtained by (a) the TDS measurement evaluated by the optimization method and (b) the 4PP measurement. The white and black spaces in the plot of the 4PP-measurement are missing data points possibly because the measuring tip did not contact the wafer correctly. This is an examplarily chosen wafer, which is in the resistivity range where the optimization method showed worse agreement with the 4PP-resistivity values. Still, the absolute value of the mean of the resistivity from the two compared methods only differ slightly by 0.8$\times$10$^{-3}$\,$\Omega$cm. The overall structure of the wafer being mostly homogeneous, but with slight variations of the resistivity close to the edges, agrees well between both methods. Again, the colorbar between (a) and (b) spans across the same resistivity range, but it is shifted by the offset of the mean of both methods for better comparability of the lateral resistivity distribution.

These two examples show the feasibility of imaging the resistivity and its distribution in a silicon wafer. The advantage of the TDS method as a high-resolution and non-contact measurement technique for possible applications for quality control is evident, since it allows to quickly detect unwanted inhomogeneities in the doping and therefore resistivity distribution of the wafers under test.

\section{Conclusion and outlook}
In conclusion, terahertz time-domain spectroscopy measurements of doped silicon wafers covering a wide range of resistivities were conducted and evaluated using the Drude model in two different ways: Once analytically by utilizing the peak-to-peak amplitude and only one frequency and once by a simulation and optimization approach based on Rouard's method making use of the whole spectrum. The results were compared to 4PP-measurements of the same samples. The results show that TDS measurements allow to determine the resistivity of doped silicon in a range from around 10$^{-3}$\,$\Omega$cm to 10$^{2}$\,$\Omega$cm well compared to the established 4PP-method. Furthermore, X-Y-scan measurements with a lateral resolution of 1\,mm of the samples allow for imaging of the spatial resistivity distribution which can be useful to detect inhomogeneities, being beneficial e.g. in the production process of the wafers for quality control. With these results together with the recently shown ability of TDS to allow for fast measurements and being a non-destructive and even non-contact measurement method, terahertz time-domain spectroscopy can possibly become a useful tool in the semiconductor industry in the future. 

\begin{backmatter}
\bmsection{Funding}
This project is supported by the Federal Ministry for Economic Affairs and Climate Action (BMWK) on the basis of a decision by the German Bundestag.


\bmsection{Disclosures}
The authors declare no conflicts of interest.

\bmsection{Data Availability Statement}
Data underlying the results presented in this paper are not publicly available at this time but may be obtained from the authors upon reasonable request.

\end{backmatter}


\bibliography{bib}

\end{document}